# Comet 81/P Wild 2: changes in the spin axis orientation during the last five apparitions

Virginio Oldani[a], Federico Manzini[a*], Paolo Ochner[b,c], Andrea Reguitti[b,d], Luigi R. Bedin[b], François Kugel[e], Jean-François Soulier[f], Orhan Erece[g,h], Doğan Tekay Köseoğlu[g,h], Çağlayan Nehir[g], Tuncay Özişik[g]

[a] *Stazione Astronomica di Sozzago, Cascina Guascona, I-28060 Sozzago (Novara), Italy*
[b] *INAF-Osservatorio Astronomico di Padova, Vicolo dell'Osservatorio 5, I-35122 Padova, Italy*
[c] *Department of Physics and Astronomy, University of Padova, Via F. Marzolo 8, I-35131 Padova, Italy*
[d] *INAF – Osservatorio Astronomico di Brera, Via E. Bianchi 46, 23807, Merate (LC), Italy*
[e] *Observatoire Chante-Perdrix, Dauban, France*
[f] *Observatoire M53 Mayenne Astronomie, Maisoncelles-Du-Maine, France*
[g] *TÜBITAK National Observatory, Akdeniz University Campus, Antalya, 07058, Türkiye*
[h] *Department of Space Sciences and Technologies, Akdeniz University, Campus, Antalya, 07058, Türkiye*

*Corresponding author
E-mail address:* manzini.ff@aruba.it

**Abstract**

Comet 81P/Wild 2 is characterized by the presence of a prominent fan-shaped dust emission originating from an active source at high latitude on the nucleus, whose axis is assumed to coincide with the comet's rotation axis. Therefore, several authors estimated the spin axis orientation of 81P in past apparitions based on the polar jet model.

By measuring the PAs of the fan on CCD images taken with different telescopes during the 2009-10 and 2022-23 apparitions, we estimated a position of the comet's spin axis at RA=295.0°±7.5°, Dec=14.5°±4.0° for the 2009-10 apparition and at RA=296.7±2.0°, Dec=17.3±2.5° for the 2022-23 apparition. Despite some degree of uncertainty of the estimate for the 2009-10 apparition, we interpolated the estimate for 2009-10 and 2022-23 with the published data of the previous apparition of 1997, to assess the presence and the extent of a drift of the pole since the 1997 passage.

The analysis over a long time span of five consecutive apparitions confirms previous observations that the spin axis of comet 81P is subject to a slow drift with variable rate, probably connected to outgassing-induced jet forces and the related non-gravitational perturbations of its orbital period.

**Keywords:** comet, 81P/Wild 2, spin axis, polar jet

## 1. Introduction

Comet 81P/Wild 2 is known as a fresh periodic comet, orbiting between Jupiter and Mars. Studies conducted by **Wild & Marsden (1978)** and **Nakano (1979)** showed that in 1974 the comet passed within just 0.0061 AU of Jupiter. The gravitational action exerted by the planet modified the comet's orbit, and its orbital period changed from approximately 43 years to 6.4 years, while its perihelion distance decreased from 4.9 AU to 1.59 AU (Small-Body Database Lookup (nasa.gov)). The Swiss astronomer Paul Wild discovered the comet on January 6, 1978, using a 40-cm Schmidt telescope at Zimmerwald, Switzerland, during its first passage on the new orbit.

It appears that **O'Meara's (1979)** brief remarks on jets and a fan in March and April 1978 are the very first description of a discrete coma morphology of 81P by a visual observer. Indeed, two major dust-emitting sources on the nucleus were identified on a set of high-resolution images taken by **Schulz et al. (2003)** at the European Southern Observatory (ESO) at La Silla, Chile, between September 1996 and April 1997: one, near the rotation pole at a latitude of about 82°; the other, on the opposite hemisphere at about 25° from the equatorial plane.

The first attempt to determine the orientation of the comet's spin axis was made by **Sekanina (2003)** on the ESO images by applying the polar jet model (**Sekanina 1979, 1981**) assuming that the projected direction of the spin axis coincided essentially with the axis of the polar dust fan, with resulting values of Right Ascension (RA) = 297°, and Declination (Dec) = -5°. The PA of the polar fan of comet 81P were measured during the same apparition by **Farnham and Schleicher (2005),** who determined the pole orientation at RA= 281°±5°, Dec= +13°±7° on CCD images taken between February and April 1997. **Vasundhara and Chakraborty (2004)** modelled the trajectories of dust grains ejected from distributed sources on the comet to explain the morphology of the fans on images taken on May 15, 1997, at the Vainu Bappu Observatory (Kavalur, India), and on the images published by Schulz et al. (2003). They derived the pole position at RA= 297° ± 5° and Dec= -10° ± 5°. The primary source was positioned at +80° ± 5° on the comet nucleus.

Comet 81P/Wild 2 was visited by the Stardust probe (NASA), on January 2, 2004. Stardust flew through the coma on the sunward side of the nucleus and missed the nucleus by just 240 kilometers. Images taken by the Stardust mission during its flyby show the comet to be a 5-kilometer oblate body. The comet's shortest axis, inferred to be the axis of rotation, was estimated at RA= 290° and Dec= +13° (**Brownlee et al. 2004, Duxbury et al. (2004)**. Based on the Stardust encounter data, **Sekanina et al. (2004)** modeled a new spin axis orientation at RA= 295°, Dec= +15°, quite far from his previous determination. However, both studies were based on the nucleus shape, as the polar jet was inactive at that time.

**Szutowicz et al. (2008)** derived a spin axis orientation at RA=290.8°, Dec=+9.7° from numerical fitting of the non-gravitational acceleration model to positional observations of the comet on two apparitions (1997, 2003). The authors also indicated possible time variations of the spin axis of the comet during five revolutions between 1978 and 2003, probably correlated with time variations of the non-gravitational perturbation of the orbital period. **Chesley and Yeomans (2005)** found a solution for the comet's pole, at RA=320°, Dec=+15° with the rotating jet model (RJM); and a different direction at RA=342°, Dec=+20° with the Extended Standard Model (ESM), based on nongravitational accelerations due to the sublimating ice.

Depending on the method used, there are large variations between the published estimated positions, which are spread within an area more than 40 degrees-wide (**Table 1**). The aim of our work was therefore to verify, by measuring the PA of the near-polar fan visible in the inner coma of comet 81P during the 2022-23 apparition, whether the polar jet model provided reliable results and whether the position of the comet's spin axis estimated with this model was comparable with those of the previous passages. We could not find published data on the orientation of the spin axis for the 2009-10 apparition, which was quite favorable for the Earth-based observations. Therefore, given the availability of CCD images of the comet for that apparition, we also conducted a retrospective analysis to estimate the orientation of the spin axis at that time and to assess the hypothesis of a possible drift of the pole since the 1997 passage.

## 2. Methods

### 2.1 CCD images and reduction

We made our analyses on images taken with different telescopes on twelve dates between October 30, 2022, and March 24, 2023, and twelve dates between January 19 and May 7, 2010 (**Table 2**). After calibration with bias, dark and flat-field, the images of the same session were aligned on the comet's nucleus and stacked in order to obtain the highest signal-to-noise ratio.

### 2.2 Measurements and data analysis

Knowing that the main feature of comet 81P is a prominent near-polar fan allowed us to determine the orientation of the pole using the same technique used successfully for comet 19P/Borrelly (**Oldani et al. 2023**): in brief, assuming that the center of the fan represents the projected spin axis, we measured its apparent PA on each date by sampling photometric profiles at various distances in pixels from the optocenter on the polar-transformed images previously processed with the Larson-Sekanina filter ($\alpha=30°$), and taking the peak brightness as a measure of the central PA of the fan.

However, differently from comet 19P, that showed a jet in a virtually polar position, some minor complications were involved with the application of this technique to comet 81P, which could introduce an uncertainty of few degrees in the measurements of the PA:

1) the main source is not centered on the pole of the nucleus, but at a lower latitude of about 81°, so that the dust emission does not go straightforward, but rather shows a corkscrew morphology that forms a broad fan-shaped structure. Thus, we would expect with rotation to observe the peak brightness of the two sides of the fan (i.e., the two edges of the emission cone) at ±9° on each side of the spin axis. In addition, since the active area is extended (it has been estimated at 4.5 km$^2$ by Sekanina (2003)), then the identification of the true axis of the cone may not always be easy.

2) The two sides of the fan on 81P showed strong curvature at different distances from the nucleus, which could also affect the position of the central axis of the fan with respect to the comet's spin axis and make its measurement inaccurate.

3) The collected images had different resolutions (above 1,000 km/pixel for the low-resolution images, see Table 2). Considering that the smallest measurement unit was one pixel, on the lower-resolution images the peak brightness could appear spread over more than a single pixel, which could have affected the accuracy of the measurement of the actual PA of the axis of the fan.

To minimize those problems and standardize the analysis, low-resolution images were upscaled 2x, then the PAs were measured on all images at fixed distances of 20, 25 and 30 pixels from the optocenter; this allowed to avoid taking measurements on the curvature of the jet and to determine its true PA with an estimated accuracy of about ±3°. The availability of a sufficient number of images collected over a time span of two months before and after perihelion also helped in that sense.

We then performed a search of pole positions within a grid of 40°x40° comprised between RA 280° - 320° and Dec -15° - +25°, that included all the published RA and Dec coordinates of the previous passages of 1997 and 2004, with the only exception of those by Chesley and Yeomans (2005), based on the ESM, that appeared too far. Using a least-squares fit, we searched for the position of the comet's pole that returned computed PAs of the spin axis that best fitted our measurements.

Finally, we run for each date a numerical modeling of the near-polar jet using a proprietary software (*Fase 6*), specifically designed to provide a graphical representation of Earth-based observations of the dust coma structures by means of the rotating jet model (details on the modeling process can be found in **Manzini et al., 2021**). The models were drawn by applying the spin axis orientation parameters (PA and angle to the sky plane) computed with the new estimated coordinates, with the jet emission located at a latitude of 81°, a rotation period of 13.5 hours (**Mueller et al. 2010**) and introducing dust parameters consistent with those published by **Farnham & Schleicher (2005)**: size 2 μm, density 0.45 g·cm⁻³ emission velocity 0.14 km·sec⁻¹.

The models were supported by a virtual representation of the comet's nucleus (supposed spherical) for each of the observing dates introducing the new estimated pole coordinates in the software *Starry Night Pro v. 8.1* (Simulation Curriculum Corp., Minnetonka, MN, USA; www.simulationcurriculum.com) to verify the exactness of the calculated positions of the comet's pole and of its insolation conditions during the observation period.

## 3. Results

### 3.1 2022-23 apparition

The observing conditions of this apparition were only fairly favorable, with the comet being at a distance from Earth between 2.6 and 1.4 AU during the observation period (**Table 2**). The way the comet appeared on the CCD images changed in relation to its relative positions with respect to the Earth and the Sun. The orbital plane of comet 81P has an inclination of only 3°; therefore, the bidimensional view of the ground-based CCD images was not or only slightly affected by changing perspectives in that sense throughout the observation period. The Earth-comet observing geometry changed to a limited extent from October 30 until perihelion on December 15, 2022, with a +33° shift in the ecliptic longitude. More significant changes occurred post-perihelion, with the ecliptic longitude changing +55° from January to March. The position of the Sun in relation to the comet as seen from Earth showed minor changes throughout the whole observation period (PA from 114° to 92°, Sun-Target-Observer angle (STO) from 24° to 34°).

The broad polar fan was well visible in the CCD images until the end of January (**Figure 1a-c**), but thereafter it appeared gradually fainter until it was no longer visible in the March 24 image (**Figure 1d**). This suggests that the comet's pole was fully insolated until the end of January, but then it gradually moved in opposite direction to the Sun and the insolation of the near-polar active area decreased until it reached a full night state in March.

The measured values of the PAs of the polar jet at 20 and 25 pixels, and between 25 and 30 pixels (as described in Section. 2.2), showed only minor differences (between 0° and 5°, mean ± S.D.: 2.2° ± 2.1° and 2.0° ± 1.8°, respectively), thus confirming a good straightness of the jet in this distance interval. The measurements at 30 pixels were chosen for the purpose of the analysis as those that were less affected by potential measurement errors in the low-resolution images. The PA was not measured on the last date of March 24 as the primary source was no longer active and the fan was not visible.

**Figure 2a** shows the best fitting solution of the pole orientation at RA = 296.7°±2.0°; Dec = 17.3°±2.5°, with values of the O-C (observed PAs – computed PAs) ≤ 2° in eight out of eleven dates and a total RMS error of only 0.54°. In **Table 3** the measured PAs are listed for each date, together with the PAs and the angles to the sky plane computed according to the found pole coordinates. Using these computed positions in the modeling software returned reproductions of the fan morphology well matching with the appearance on the CCD images (**Figure 3 top panel**).

Furthermore, using the estimated pole coordinates to verify the extent of the insolation of the comet's pole throughout the observation period confirmed that it moved to a night-day state in February and in full night in March, consistently with the observation of the gradual fading and subsequent disappearance of the fan (**Figure 4**). We also verified whether the source would have moved into full night earlier in 2023 if the 1997 pole coordinates were used and could confirm that it would have been in full night already around January 20, i.e. more than a month earlier, while according to the new estimated position of the pole it was still fully illuminated on that date. This observation supports the fact that that the pole has actually moved since then.

The new coordinates are well within the predefined grid of 40° comprising the published positions, at a long distance from the estimate by **Farnham and Schleicher (2005)** and at a shorter distance from the estimates by **Sekanina et al. (2004), Brownlee et al. (2004) and Duxbury et al. (2004)**, suggesting that the position of the pole has been subject to a significant migration compared to the previous comet's passages of 1997 and 2004.

## 3.2  2010 apparition

The observing conditions of this apparition were quite favorable, with the comet reaching the minimum distance from Earth of 0.68 AU on April 5. The Earth-comet observing geometry remained relatively constant throughout the whole observation period from end December 2009 to early May 2010, with a maximum variation of +30° in the ecliptic longitude. Similarly, the position of the Sun in relation to the comet as seen from Earth showed little changes before perihelion (PA from 113° to 108°, STO from 35° to 30°), but it changed significantly post-perihelion (PA moving clockwise from 108° to 315° and STO from 30° to 5° in April, to 10° on May 7).

The main fan was well visible in the CCD images until the end of March, but after then it appeared gradually fainter until it was no longer visible in the image of May 7, suggesting that the comet's pole had moved into the night side (**Figure 5**). At that time, the southern hemisphere of the nucleus should have been fully insolated, and in fact a new sunward pointing jet appeared on this side, probably related to the known source located at a mid-southern latitude (**Sekanina, 2004; Farnham and Schleicher, 2005**).

The differences between the measured values of the PAs of the polar jet at 20 and 25 pixels, and between 25 and 30 pixels, were small (< 3°, mean ± S.D.: 0.5°±1.0° and -0.3°±1.6°, respectively), thus confirming a good straightness of the jet in this distance interval. Nine out of twelve images were taken with large telescopes and have high resolution ranging between 120 and 180 km·pixel[-1], which allowed high-precision measurements (**Table 3**). Despite this, the least-squares fit did not provide a reliable solution: the minimization process produced a good fit only until March 5, with the best estimate of the polar coordinates at RA=295°±7.5°, Dec=14.5°±4.0° (RMSE=1.2°). This position however resulted in increasing deviations (from 5° to 22°) between the measured PA of the fan and the computed PA of the spin axis on the subsequent observation dates (**Figure 2b**). No acceptable solution was found when all the measured PAs were considered in the fitting process, suggesting that there was something wrong with the measurements taken on the images of those dates.

Interestingly, starting from mid-March, the main fan showed a progressive clockwise rotation towards North. In addition, a second, weaker fan appeared, just south of the main fan and almost parallel to it (**Figure 5b and 5c**). This second fan was initially collimated and well separated, but then widened until it merged into the main fan, making it difficult to identify its true axis with certainty. The axis of the second fan was instead consistent with the PA of the spin axis computed according to the estimated pole coordinates, suggesting that it had caused a displacement of the main fan and a progressive dissociation of its apparent PA from the predicted PA of the spin axis. This finding was confirmed by the contemporary appearance of a pair of jets originating from two diverse sources almost symmetrical to each other at mid-low latitudes on each side of the nucleus equator; the apparent axes of their emission cones were coincident with the computed PA of the spin axis (**Figure 5c**). Although these findings could possibly explain of the observed discrepancies between the measured PAs of the main fan and the computed PAs of the spin axis in the second half of the observation period, this issue has a number of important implications, which are discussed more fully in Section 4.

As a further verification of this anomaly, the results of the numerical modeling of the fan in the dates following March 5, obtained by entering the measured PAs of the spin axis, were compared with those obtained by entering the PA computed from the estimated coordinates of the pole. The modeling returned a better match of the shape and position of the main fan with that of the CCD images in the first case (**Figure 3 bottom panel**), but conversely it showed a good representation of the second fan in the second case. These results confirmed the observed inconsistency between the apparent PA of the (center of the) main fan and the true PA of the spin axis.

The evaluation of the evolution of the illumination conditions of the near-polar active area during the observation period using the estimated polar coordinates confirmed the observation of the disappearance of the main fan in May, as well as of the appearance of the meridional emission due to the contemporary activation of the source located on the southern hemisphere of the nucleus (**Figure 6**). We also verified whether the main polar source would have moved into full night earlier in 2010 if the pole coordinates had not changed since 1997, and we could see that it would have been in full night already around March 21, i.e., about a month earlier. These observations support the fact that the estimated coordinates for 2010 are reasonably correct, and that the pole had indeed migrated from its 1997 position.

### 3.3 Analysis of the drift of the spin axis over five orbits in the period 1997-2023.

Following the finding of a large difference between the 1997 and 2023 positions of the comet's pole, we interpolated the available data for the apparition of 1997 by **Farnham & Schleicher (2005)** with our estimates for 2023 and 2010 (despite the uncertainty of this latter estimate), assuming a regular drift of the comet's spin axis over the last five apparitions. We only included these positions because they were well determined from the same technique; all the other estimates were excluded from the interpolation because they were derived with different methods or averaged over multiple orbits.

Treating the pole solutions as exact determinations, the curve passing through each point shows a perfect fit ($R^2$=1.0) with a parabolic shape; however, considering the uncertainties in the estimated positions (particularly for 2010), also a linear regression would reasonably fit ($R^2$=0.68) (**Figure 7**). Interestingly, the positions estimated by **Brownlee et al. (2004), Duxbury et al. (2004)** and **Sekanina et al. (2004)** for the 2004 passage, as well as that by **Szutowicz et al. (2008)**, based on the two passages of 1997 and 2003, all fall in intermediate positions close to the fitting curve, which confirms that they were correct. It looks however that the solutions for the comet passages considered in this analysis do not lie at regular intervals, suggesting that the rate of the pole drift varied between each orbit.

### 4. Discussion

The estimate of the comet's pole coordinates resulting from the best fit of the measured PAs of the main fan appears quite good for the 2022-23 apparition, while it shows some degree of uncertainty in the second part of the observation period of the 2010 apparition.

Small errors in the measurement of the PA of the fan on some dates cannot be totally ruled out, given the potential complications of the method, as described in section 2.2. Nonetheless, we think that possible measurement errors did not affect the estimation of the pole coordinates significantly. The gradual fading and subsequent disappearance of the fan observed on the images of the last dates of both apparitions was instrumental to confirm the evolution of the extent of insolation of the active source of the primary fan predicted by our estimates of the orientation of the spin axis and, consequently, their correctness. Likewise, the finding that the active source would have been in full night a month earlier in both 2010 and 2023 if the pole coordinates estimated in 1997 had been unchanged in these last passages, also supports the fact that our estimates are reasonably correct.

One potential cause of the mismatches between the computed PA of the spin axis and the observed PA of the fan in March-April 2010 could be related to the changes in the orientation of the spin axis and of the location of the active source with respect to the Sun. According to the estimated coordinates, the comet's north pole moved in opposite direction to the Sun in March and April 2010, and the intensity of the insolation of the source region at 81° latitude on the nucleus rapidly decreased as the angle between the sub-solar latitude and the surface normal to the active region increased from 75° in March to 104° on May 7, when the fan was turned off (**Figure 5d and Figure 6**). In these conditions, not only were the outgassing activity and the shape of the fan strongly affected by the grazing and intermittent illumination, but also the projected PA of the fan could appear shifted relative to the true PA of the spin axis to an observer on Earth.

The intermittent insolation of the active area could also have determined a change in the morphology of the fan as an effect of the thermal lag, showing only the portion of the emission cone exposed to daytime insolation (**Sekanina 1987**) . It is thus possible that the PA of the emission forming this partial cone appeared displaced towards the day side of the nucleus with respect to the direction of the spin axis.

Additional effects of shape and topography of the nucleus that our spherical model neglects may also have played a role. Our assumption that the jet is emitted normal to a spherical surface may certainly be a source of error: comets are known to frequently have jets associated with pits and slopes (and 81P has a very rough surface with lots of topographic features), so the emission is not normal to the surface and can introduce a systematic offset between the fan center and the pole axis. As the sub-solar latitude changes with time, the magnitude of this offset can also change. Furthermore, a slope changes the illumination conditions, which can induce variations of the activity with time. These effects have the potential to shift the measured pole position by several degrees.

Another potential cause that led us to misinterpret the PAs of the main fan PAs could be the occurrence of an outburst that ejected substantial amounts of dust and displaced the apparent center of the fan. In fact, **Bertini et al (2012)** observed two increases in activity, peaking around 60 and 160 days after perihelion. Amateur data collected at close time intervals over a longer period suggest that the increase in dust production had started earlier, before the end of March (see for example the light curve of the comet at http://aerith.net/comet/catalog /0081P/2010.html). The appearance of the second fan just south of the main one in mid-March supports the hypothesis that the outburst occurred at a latitude on the nucleus not far from the source of the main fan, causing its axis to shift northward. This may have misled our measurements of the PA of the rotation axis, which in reality had not changed. The fact that the projected axes of the second near-polar fan and of the pair of "near-equatorial" jets that appeared from mid-March, as well as the position of the southern jet that appeared in May, were all consistent with the computed PAs of the spin axis for those dates, also favors this hypothesis.

It is possible that this latter is the most plausible explanation, although all the above hypotheses may have contributed to altering our perception of what was observed on CCD images and, consequently, the measurements of the PAs. What appears undeniable, however, is that a series of circumstances occurred in 2010 (but not in 2023) that profoundly affected the measurements. On the other hand, the large variability in the positions estimated by other authors who adopted the polar jet model (**Sekanina 2003; Vasundhara and Chakraborty 2004; Farnham and Schleicher 2005; Chesley & Yeomans 2005**), do confirm the intrinsic difficulty of performing these measurements when the real conditions on the comet's nucleus are different from what is expected from theory.

It must be said that the insolation conditions of the active region in 2022-23 were similar to those observed in 2010, but no shift of the PA of the fan with respect to the PA of the spin axis was detected, nor the appearance of the second fan or of the near-equatorial jets. It should however be considered that the resolution of the images in 2023 was generally much lower because the comet was at a greater distance from Earth, so such features may not have been appreciable, or simply that some of the phenomena described (e.g. an outburst) did not occur.

All considered, we decided to accept our estimate of the position of the pole of 81P in 2010 as sufficiently reliable, despite the high degree of uncertainty and the fact that it was based only on the measurements of the first half of the observation period.

Unfortunately, our estimate for 2010 cannot be compared with other estimates for that apparition, as there are no published data. **Lin et al. (2012)** performed a morphological analysis of the coma of comet 81P on twelve images taken at the Lulin Observatory in Taiwan between January 14 and August 1, 2010. They identified up to seven different dust structures: in particular, the broad northeast jet was divided into two substructures (named A and B). The authors measured the PAs of the two substructures, while in our interpretation the same structures appear as being the two sides of the emission cone of the fan. **Bertini et al. (2012)** analyzed the dust structures in the coma of comet 81P on images acquired at the 2.56-m NOT telescope in La Palma, and at the 0.82-m IAC telescope in Tenerife during the perihelion passage of 2010. The authors had a similar interpretation as **Lin et al. (2012)**, although they stated that most of the identified features could be the same structures originated from the same source. Actually, in few cases the measured PAs of these structures were close to our measurements performed on the same dates despite they had been interpreted differently.

To assess whether there has been a true migration of the pole or a precessional movement since 1997, we chose to proceed by interpolating the positions for 1997, 2010 and 2023, and then verifying whether the other published estimates lay close to or on the fitting curve. For 1997 we chose to keep the position by **Farnham & Schleicher (2005)**, also based on the polar jet model, as it has been considered more likely than the one found by **Sekanina (2003)** (**Gutiérrez & Davidsson 2007**). We instead decided not to consider the estimates for 2004, either because they were obtained with different methods (e.g., the nucleus shape model), or because they seemed less accurate and outside the predefined grid. This approach presents obvious intrinsic weaknesses but could be considered an acceptable compromise to obtain useful information on the evolution of the orientation of the spin axis of comet 81P over a time span of five orbits.

We could not include a position for the 2016 passage, as there are no published data on the orientation of the spin axis at that time. Assuming that the comet's pole was in an intermediate position between those estimated for 2010 and 2023, it should have been oriented towards the Sun from August 2015 to end July 2016 (the perihelion passage was on July 20, 2016), with the near-polar active area fully insolated. In fact, an image taken on March 18, 2016 (r=1.99 AU, delta=1.8 AU) retrieved from the archives of the 4.2-m William Herschel Telescope (La Palma, Canary Islands) shows a well visible broad fan, with its axis being at an approximate PA of 303°. This is consistent with the computed PA of the spin axis for that date if the pole had been directed to a position approximately between those of 2010 and 2023.

The orbital motion of comet 81P is known to be affected by fairly minor nongravitational perturbations, which are likely a product of the momentum transferred to the comet's mass by anisotropic outgassing from the discrete emission sources (**Sekanina, 2003**). The comet is undergoing, since 1978, a well-established decrease of its orbital period, but in addition, it seems that the rate of change of the orbital period is gradually decreasing. This indicates that this comet experiences significant sublimation induced forces and that they are changing with time (**Gutiérrez & Davidsson 2007**). The fact that the spin axis orientation appears to be different in each of the last five previous apparitions of 1997, 2004, 2010, 2016 and 2023, raises the question if both phenomena, i.e., the variation in the rate of change of the orbital period and the possible outgassing-induced drift of the spin axis orientation, are connected.

Several authors (e.g., **Samarasinha and Belton 1995; Neishtadt et al. 2002; Gutiérrez et al. 2003, 2005**) have investigated from a theoretical standpoint the effect of the outgassing-induced torque on the evolution of rotational parameters of comets. The lack of clear and definitive observational evidence of rotational evolution, as predicted by theoretical modeling, could be due to the difficulty in determining accurate cometary spin parameters. For example, **Chesley & Yeomans (2005)** derived pole coordinates from an orbital fit using the Rotating Jet Model for three apparitions (1990–1997–2003) that compare favorably with those obtained by other authors. However, they obtained quite different results using the Extended Standard Model, based on nongravitational accelerations due to the sublimating ice. Following the observation of a pre-perihelion increase in activity associated with the near-polar jet, as opposed to the presence of a mid-latitude jet in the southern hemisphere, they concluded for a substantial seasonal effect of jetting, i.e. that nongravitational thrusting acting on the nucleus takes place only when this active area is exposed to sunlight, which yields important constraints on the orientation of the nucleus spin pole.

**Szutowicz et al. (2008),** applying a two-source nongravitational spotty model, noted that the evolution of the spin axis direction is correlated with time variations of the non-gravitational perturbation of the orbital period. Orbital linkages of two and three successive apparitions of 81P (1997-2003 and 1990-1997-2003) showed an evolution of the change in the orbital period, possible time variations of the spin axis of the comet and an increase of the source areas during five revolutions from 1978 to 2003.

The rotational evolution of comet 81P and possible drift of its spin axis due to the outgassing-induced jet force was extensively studied by **Gutiérrez & Davidsson (2007)**, who found a decrease in the orbital period of 81P from 2004 of -5 to -1 min·orbit[-1], similar to the orbital period deceleration observed from 1988 up to 1997, and probably secondary to a slight drift of the spin axis orientation towards larger ecliptic longitudes. According to their results the most probable variation of the spin axis is 4°–8° per orbit.

Our assumption of a regular drift of the orientation of the spin axis during the past five orbits, which was at the basis of our approach of interpolating the estimated coordinates for the 1997, 2010 and 2023 passages, appears in contrast with the observation of a variable rate of the drift in the different passages. Indeed, looking at the trend of the curves describing the changes in the position of the pole of 81P resulting from the coordinates adopted for the five apparitions, the rate of change appears variable and decreasing from 2004 onwards, although the changes between one apparition and the following are generally small, suggesting that the variability in the rate of change is limited to a narrow range.

An interesting comparison of what was observed for comet 81P can be made with comet 19P, which also showed a migration of the pole during the last few perihelion passages, probably induced by variations in the outgassing activity of a dominant polar jet (**Oldani et al. 2023**). The variations observed in the drift rate of the pole of comet 81P in the analyzed orbits could therefore be related to variations in the activity of the polar jet as a function of the distance from the sun and of the different insolation conditions of the active area in the various perihelion passages.

In conclusion, we believe that, despite its inherent weaknesses, our approach provides a realistic picture of the evolution of the orientation of the spin axis of 81P during the last five orbits and confirms that it is subject to a slow and irregular drift or to a secular precession.

The next perihelion passage of 81P is expected on May 19, 2029; the geometric conditions of ground-based observations will not be very favorable, with a distance from the comet just under 1 AU more than two months before perihelion. In that period, however, the pole will be constantly illuminated with the consequent development of the emission activity that originates the polar fan. Based on our determination of the direction of the spin axis during the last passage in 2023, the active area that generates the fan should go into shadow (with the fan reducing its emissivity) around the end of June 2029.

## 5. Conclusions

From the analysis of CCD images of comet 81P/Wild 2 taken with different telescopes during the 2022-23 and 2010 apparitions, we found as main feature of the inner coma the near-polar fan already observed by other authors during previous perihelion passages. This finding indicates a recurrent main activity always driven by the same areas on the nucleus, producing similar coma structures in the different years.

By measuring the PAs of the fan, we estimated a position of the comet's spin axis at RA= 296.7±2.0°; Dec= 17.3±2.5° for the 2022-23 apparition and at RA= 295.0±7.5°, Dec= 14.5±4.0° for the 2009-10 apparition.

Despite some degree of uncertainty of the latter estimate, we interpolated these coordinates with the published data related to the previous apparition of 1997, to assess the presence and the extent of a drift of the pole since the 1997 passage. The analysis over a long time span of five consecutive apparitions confirms previous observations that the spin axis of comet 81P is subject to a slow drift with variable rate, probably connected with outgassing-induced jet forces and the related non-gravitational perturbations of its orbital period.

## Acknowledgements

We would like to thank:
Ivano Bertini for the useful discussion and for sharing his original images.
Yuuji Oshima for sharing his useful images.
Tenay Rambaldo-Saguner for her kind support
TÜBITAK National Observatory (Türkiye) for partial support in using the T100 telescope with Director Discretionary Time.
The reviewers for the precious suggestions that allowed to greatly improve this paper.
This work made use of publicly available images retrieved from the archives of the KISO, KPNO, Faulkes and NOT Observatory.

## FIGURES

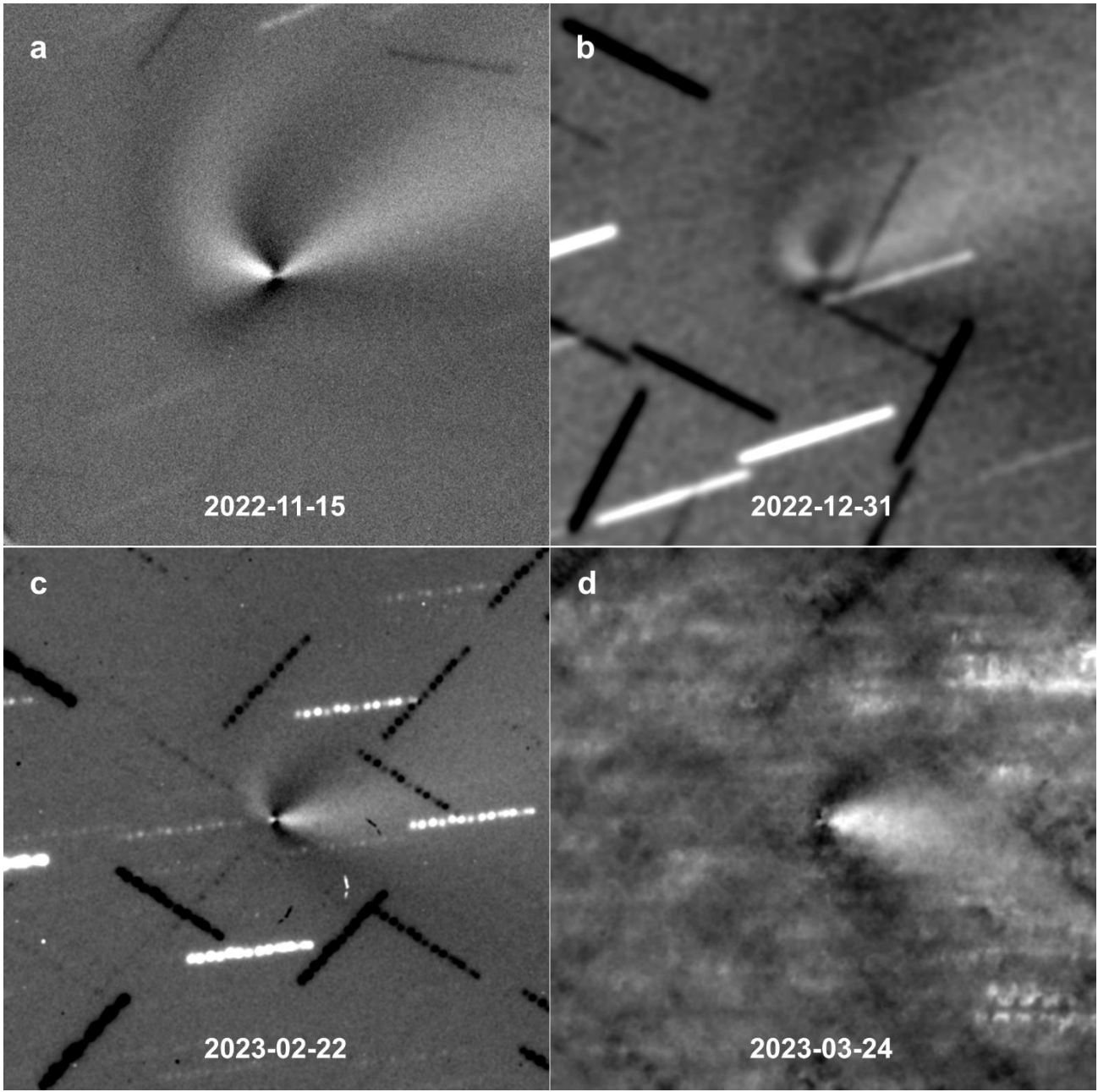

**Fig. 1**. CCD images of comet 81P taken on four dates of the 2022-23 observation period. The broad polar fan is clearly visible in the first two dates, while it appears fading on February 22, 2023, and no longer visible on March 24, when a dust emission appeared on the southern hemisphere of the nucleus. Details of the images are given in Table 2.

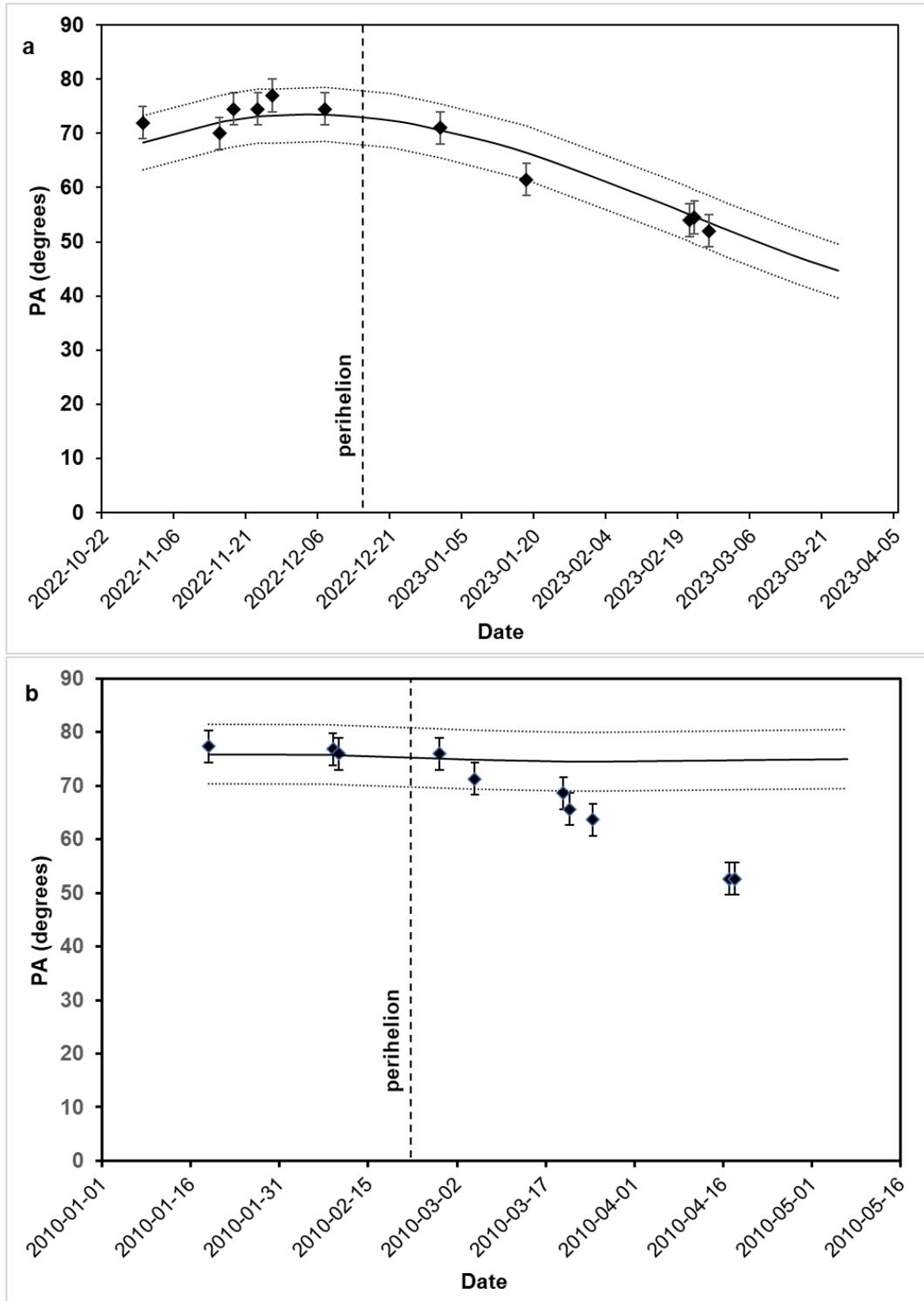

**Fig. 2**. The apparent position angle of the spin axis as a function of time in 2022-23 (panel a) and in 2010 (panel b). The diamonds (with related error bars) indicate the measured PAs of the main jet, while the solid lines represent the computed PAs for our best fitting pole solution and the dotted lines the associated 95% Confidence Interval. For 2010 the plotted pole solution is based only on the measurements until March 5; increasing deviations were observed between the measured and the expected PAs on the later observation dates, indicating measurements problems.

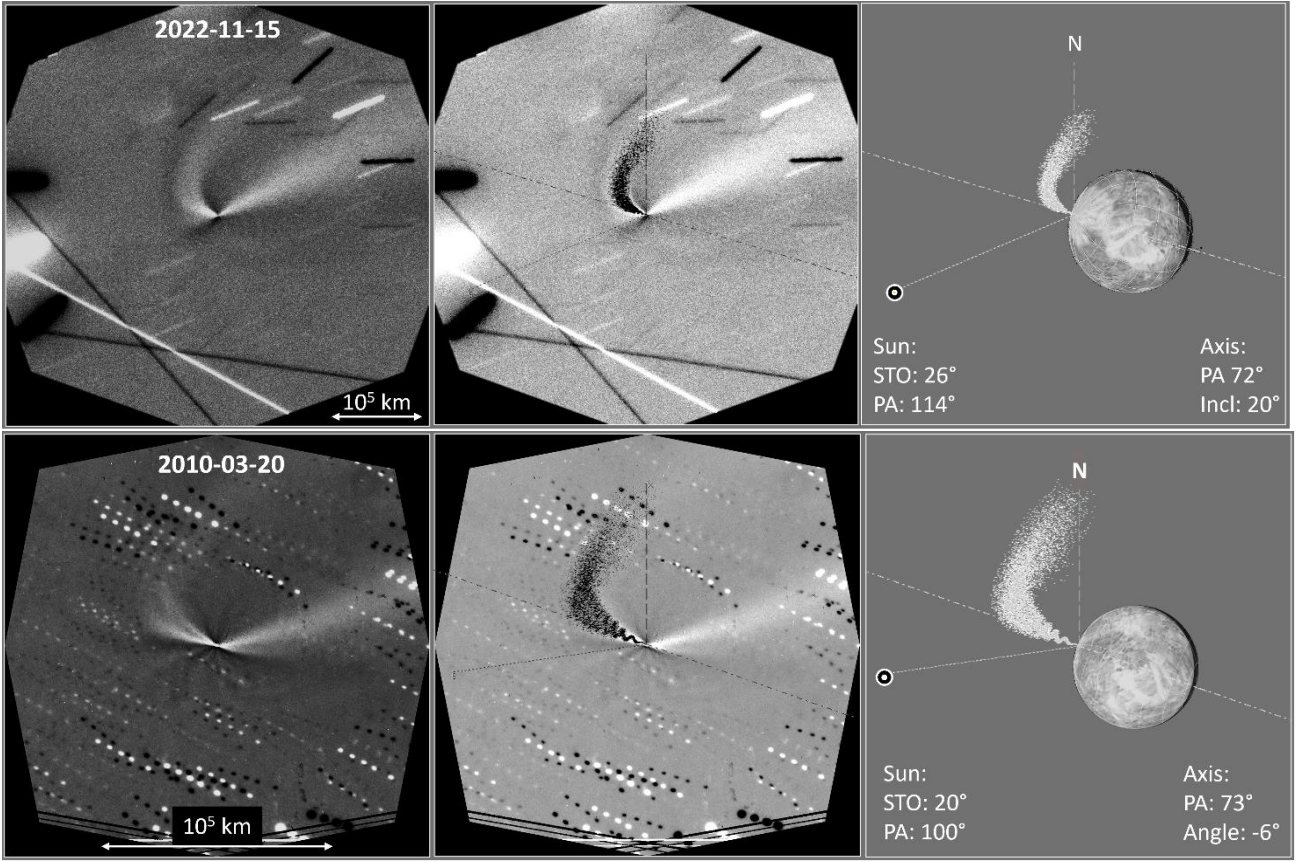

**Figure 3.** First column: CCD images taken on two dates of the 2022-23 and 2010 passages, respectively, processed with the Larson-Sekanina algorithm centered on the optocenter (alpha=10°) to highlight the polar fan. The images have different resolutions but show a similar morphology of the fan: the two sides of the emission cone are clearly visible. Second column: a numerical model of the fan (black) is superimposed to the CCD image to verify the exact correspondence using the measured PA of the fan. Third column: the model of the fan is shown on a simulated nucleus (assumed spherical, not in scale) to reproduce the spin axis orientation and the extent of insolation of the nucleus (large bright portion of the sphere vs thin black slice) as seen from Earth.

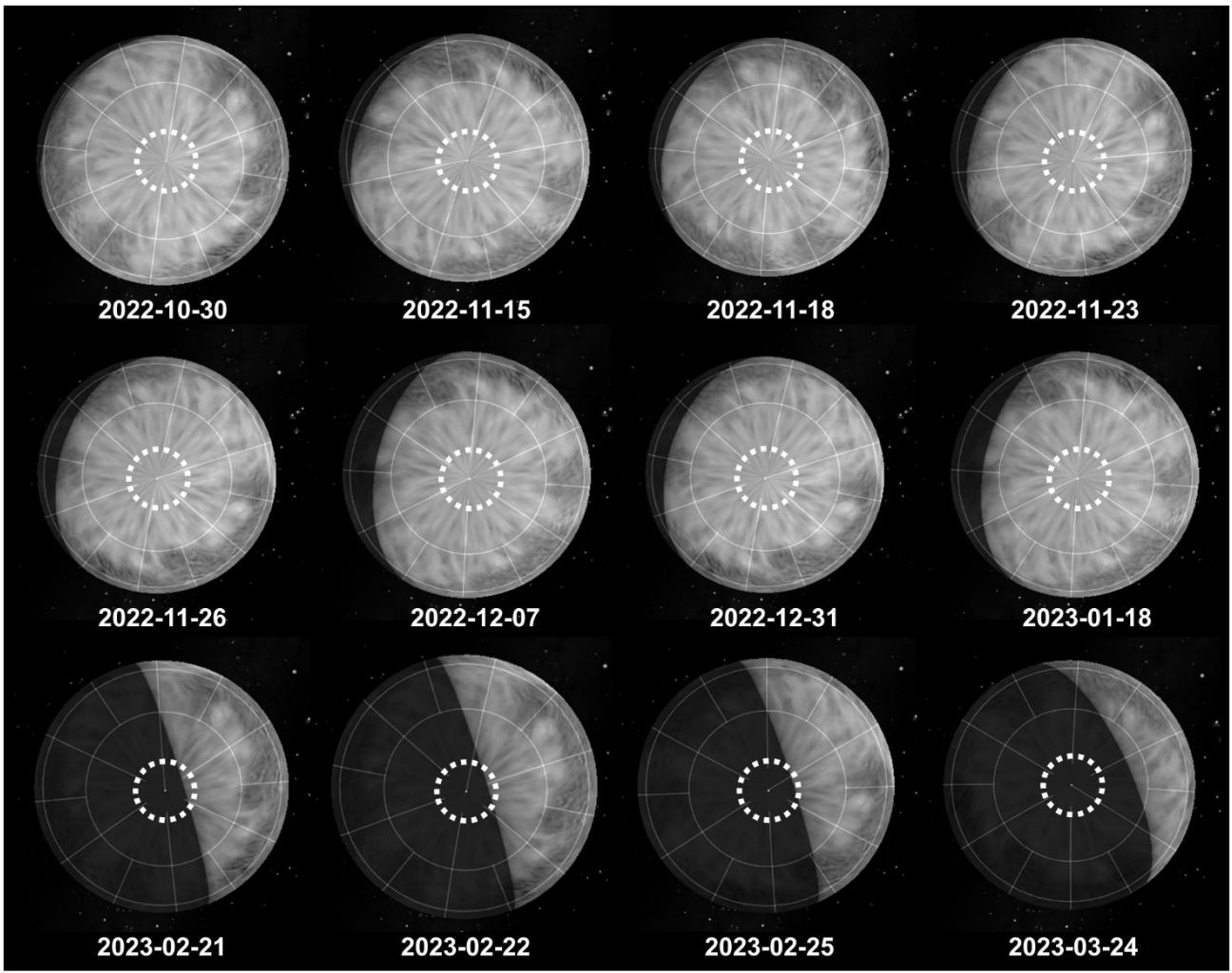

**Fig.4**. Illumination conditions of the northern polar area of comet 81P during the 2022-23 observation period, according to the estimated coordinates of the pole. The white dotted circles indicate the latitude of the main active region originating the polar fan. This was only partially insolated in February 2023 and in full night in March, when the main fan shut down. Simulations performed with the software Starry Night Pro v. 8.1.

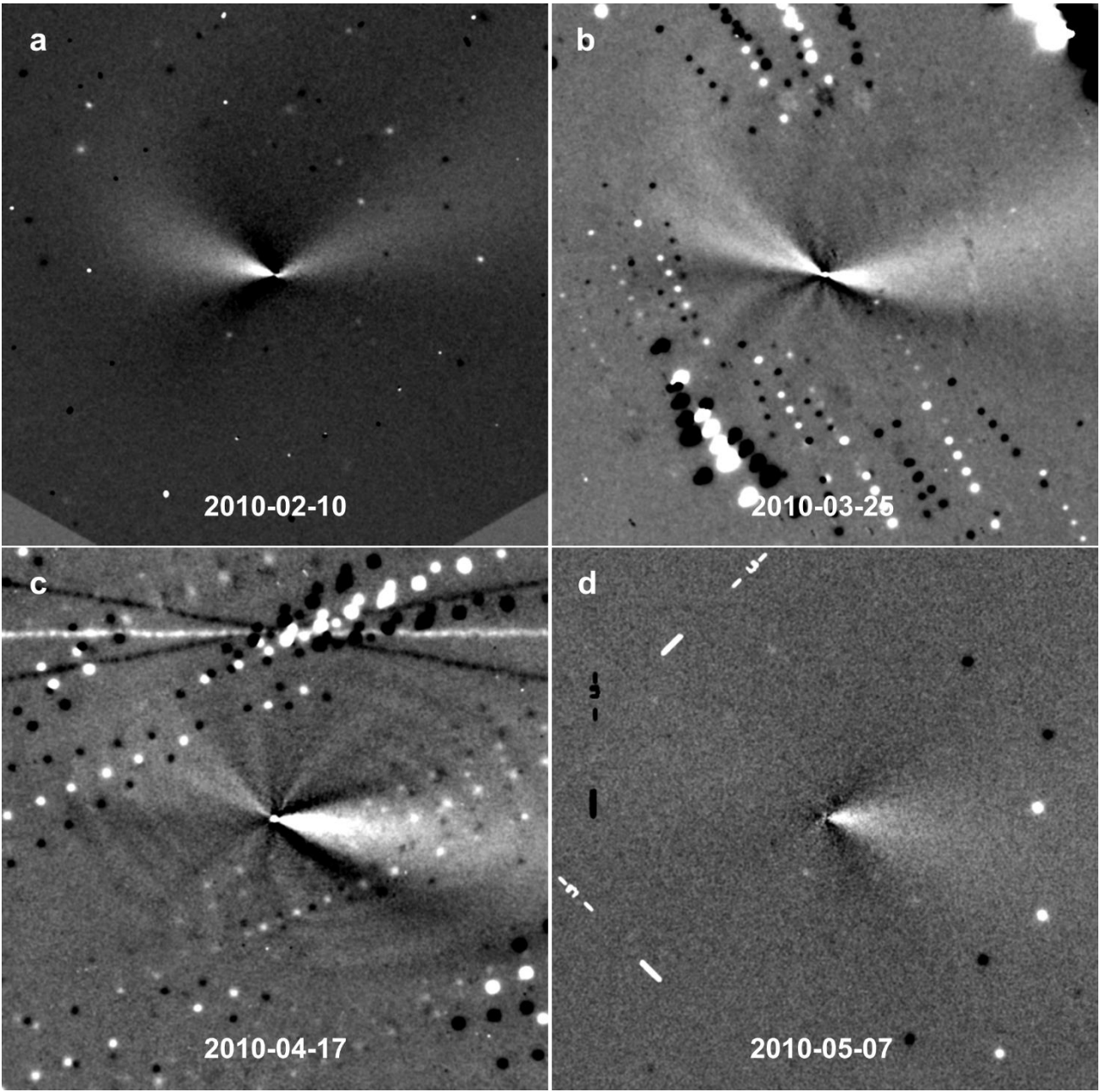

**Fig. 5.** CCD images of comet 81P taken on four dates of the 2010 observation period. The broad polar fan is clearly visible in the first two dates, while it appears almost merged with a second fan and shifted northward on April 17. On the same date also a pair of symmetrical jets appeared at each side of the nucleus equator, whose axis is coincident with that of the second fan. The polar emissions are no longer visible on March 24, suggesting that the north pole was in full night, and a new dust emission appeared on the southern hemisphere of the nucleus. Details of the images are given in Table 2.

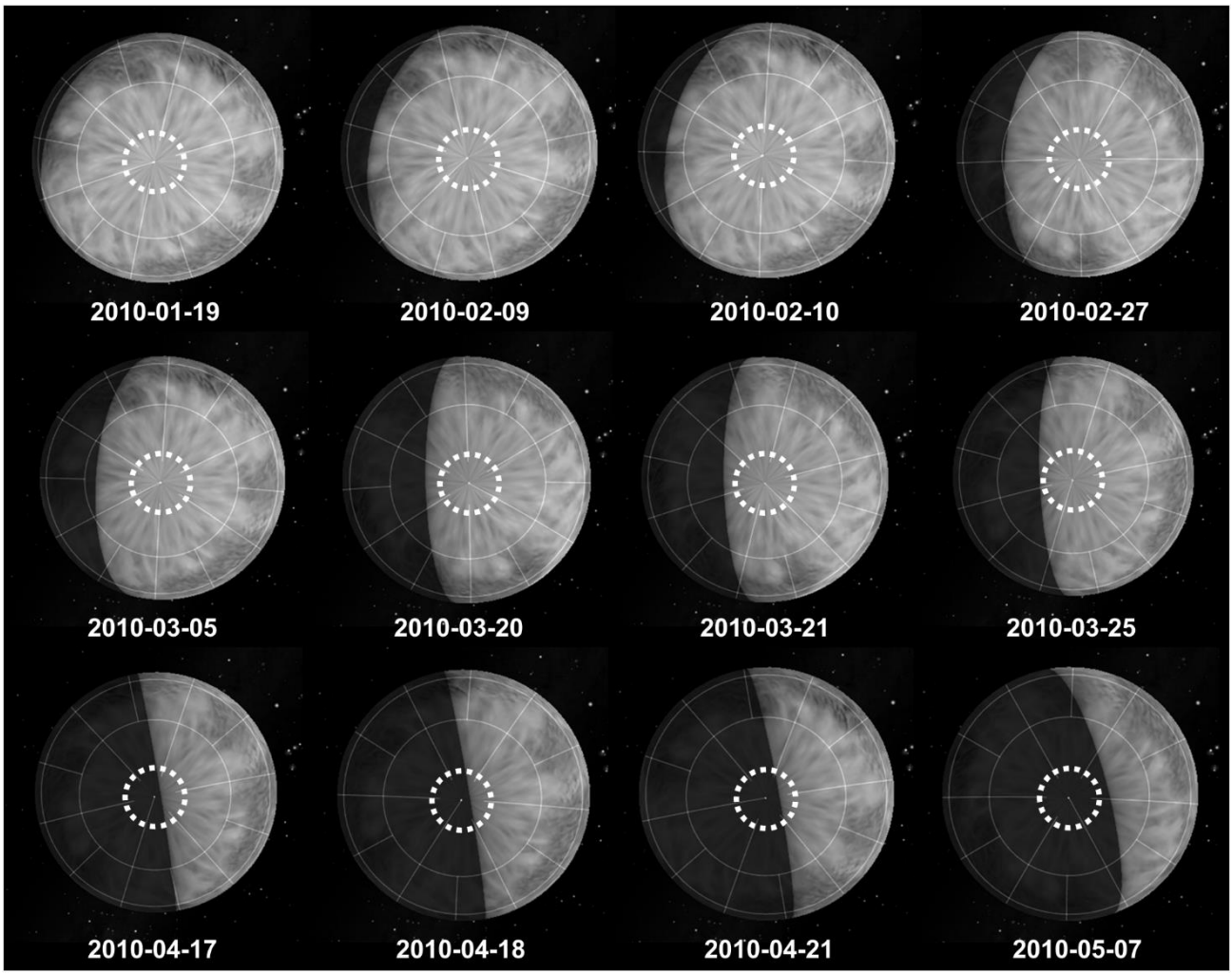

**Fig. 6.** Illumination conditions of the northern polar area of comet 81P during the 2010 observation period, according to the estimated coordinates of the pole. The white dotted circles indicate the latitude of the main active region originating the polar fan. This was only partially insolated in April 2010 and in full night in May, when the main fan shut down. Simulations performed with the software Starry Night Pro v. 8.1.

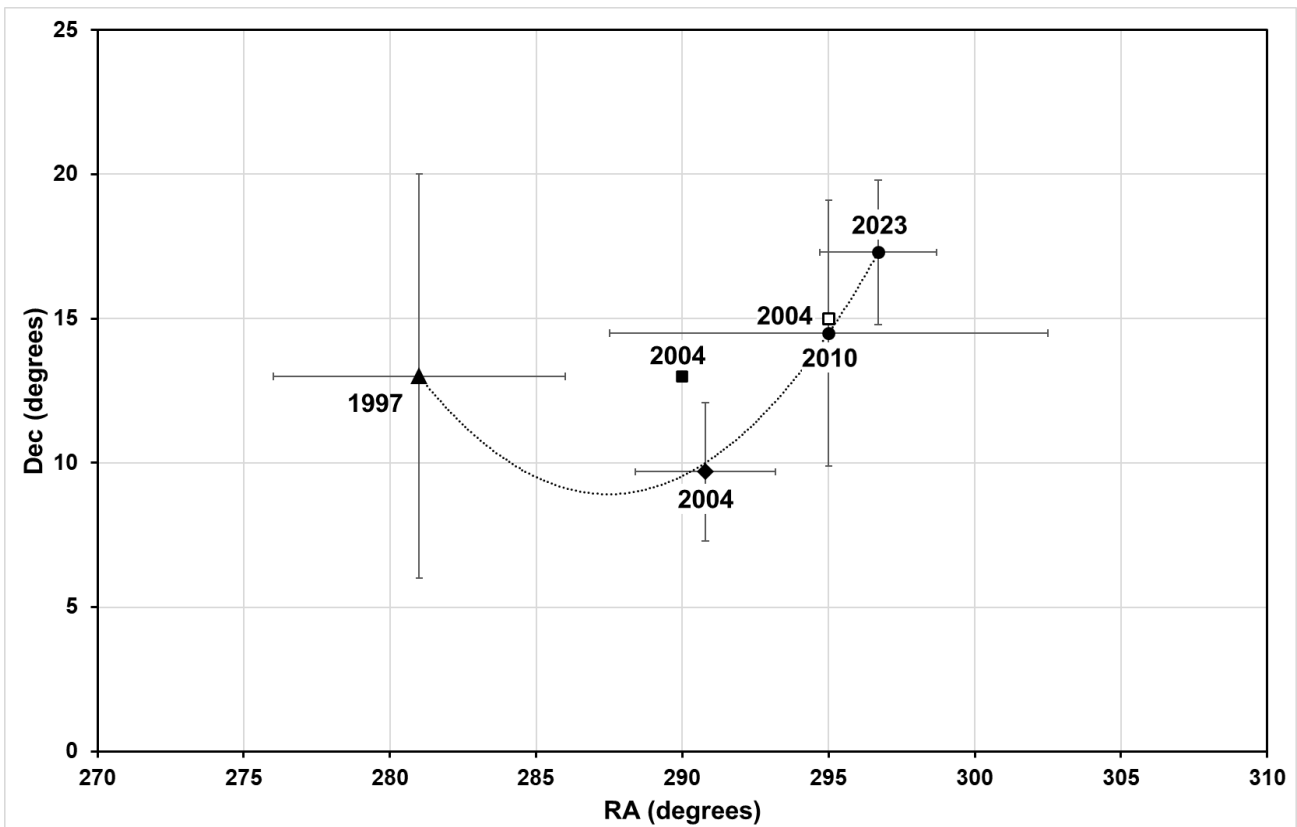

**Fig. 7.** Theoretical path of the pole drift of comet 81P from 1997 to 2023. The dotted line is the curve interpolating the positions for 1997, 2010 and 2023 only, while the positions for the 2004 passage are shown to verify whether they lie in intermediate positions on the curve or close to it. The positions are shown with the associated error where available. Filled triangle: Farnham & Schleicher (2005). Filled box: Duxbury and Newburn (2004) and Brownlee et al. (2004). Filled diamond: Szutowicz et al. (2008). Empty box: Sekanina et al. (2004). Filled circles: this paper.

**TABLES**

| AUTHOR | RA (°) | Dec (°) | Method | Year(s) | N. Appar. |
|--------|--------|---------|--------|---------|-----------|
| Sekanina (2003) | 297 | -5 | Polar jet | 1996-7 | 1 |
| Farnham & Schleicher (2005) | 281±5 | +13±7 | Polar jet | 1997 | 1 |
| Vasundhara & Chakraborty (2004) | 297±5 | -10±5 | Polar jet | 1997 | 1 |
| Chesley & Yeomans (2005) | 320 | +15 | Polar jet | 1988-2003 | 3 |
| Chesley & Yeomans (2005) | 342 | +20 | Non gravitational | 1988-2003 | 3 |
| Szutowicz et al. (2008) | 290.8±2.4 | +9.7±2.4 | Non gravitational | 1997-2003 | 2 |
| Sekanina et al. (2004) | 295 | +15 | Nucleus shape | 2004 | Stardust |
| Duxbury and Newburn (2004) | 290 | +13.0 | Nucleus shape | 2004 | Stardust |
| Brownlee et al. (2004) | 290 | +13.0 | Nucleus shape | 2004 | Stardust |

**Table 1**. Coordinates of the pole of comet 81P estimated by various authors during previous apparitions.

| Date (UT) | MPC code – Telescope | Total exp. time (s) | Resolution (km/pixel) | Filter | Δ (AU) | r (AU) | PA Sun (°) | STO (°) |
|---|---|---|---|---|---|---|---|---|
| 2010-01-19 | 381 - 1.05 m | 350 | 577 (2x) | R | 1.06 | 1.63 | 112.4 | 35.3 |
| 2010-02-09 | Z23 - 2.56 m | n.a. | 123 | R | 0.89 | 1.60 | 110.0 | 33.2 |
| 2010-02-10 | F65 - 2.0 m | 30 | 180 | R | 0.88 | 1.60 | 109.8 | 33.0 |
| 2010-02-27 | C10 - 0.4 m | 1200 | 435 (2x) | Clear | 0.78 | 1.60 | 106.7 | 28.9 |
| 2010-03-05 | C10 - 0.4 m | 1080 | 419 (2x) | Clear | 0.75 | 1.60 | 105.3 | 26.9 |
| 2010-03-20 | 695 - 2.1 m | 3000 | 151 | R | 0.70 | 1.62 | 100.3 | 20.5 |
| 2010-03-21 | 695 - 2.1 m | 2400 | 151 | R | 0.69 | 1.62 | 99.9 | 20.0 |
| 2010-03-25 | 695 - 2.1 m | 840 | 149 | R | 0.68 | 1.63 | 97.9 | 17.9 |
| 2010-04-17 | 695 - 2.1 m | 960 | 149 | R | 0.68 | 1.68 | 56.2 | 5.6 |
| 2010-04-18 | 695 - 2.1 m | 840 | 149 | R | 0.69 | 1.69 | 50.5 | 5.3 |
| 2010-04-21 | 695 - 2.1 m | 600 | 151 | R | 0.69 | 1.70 | 29.3 | 4.6 |
| 2010-05-07 | E10 - 2.0 m | 10 | 155 | R | 0.76 | 1.75 | 314.8 | 9.8 |
| 2022-10-30 | A77 - 0.4 m | 810 | 676 (2x) | Clear | 2.26 | 1.66 | 114.0 | 23.7 |
| 2022-11-15 | 098 - 1.82 m | 1200 | 390 | R | 2.13 | 1.63 | 114.2 | 26.4 |
| 2022-11-18 | A77 - 0.4 m | 600 | 633 (2x) | Clear | 2.12 | 1.62 | 114.2 | 26.7 |
| 2022-11-23 | A12 - 0.4 m | 1500 | 632 (2x) | Clear | 2.08 | 1.61 | 114.0 | 27.6 |
| 2022-11-26 | A77 - 0.4 m | 945 | 617 (2x) | Clear | 2.06 | 1.61 | 113.9 | 27.9 |
| 2022-12-07 | A77 - 0.4 m | 315 | 596 (2x) | Clear | 1.99 | 1.60 | 113.0 | 29.3 |
| 2022-12-31 | D81 - 0.3 m | 3600 | 804 (2x) | Clear | 1.85 | 1.61 | 109.5 | 32.1 |
| 2023-01-18 | A77 - 0.4 m | 420 | 526 (2x) | Clear | 1.76 | 1.63 | 108.6 | 33.5 |
| 2023-02-21 | A84 -1.0 m | 3600 | 356 | R | 1.58 | 1.73 | 98.5 | 34.4 |
| 2023-02-22 | A84 -1.0 m | 1800 | 355 | R | 1.58 | 1.73 | 98.3 | 34.3 |
| 2023-02-25 | 098 - 1.82 m | 2700 | 285 | R | 1.56 | 1.75 | 97.6 | 34.3 |
| 2023-03-24 | A84 - 1.0 m | 3960 | 318 | R | 1.42 | 1.86 | 92.2 | 31.9 |

**Table 2**. List of images used and of the main geometry parameters. Each MPC code corresponds to an observatory (see legend below). The parameters delta, r, PA Sun, STO have been taken from the Horizon/JPL web app Horizons System (nasa.gov).

| MPC code | Observatory |
|----------|-------------|
| 381 | KISO Observatory - Nagano (Japan) |
| Z23 | Nordic Optical Telescope - La Palma (Spain) |
| F65 | Faulkes South – Siding Springs, New South Wales (Australia) |
| C10 | M53 Mayenne Astronomie, Maisoncelles-Du-Maine (France) |
| 695 | Kitt Peak National Observatory – Arizona (USA) |
| E10 | Faulkes North – Haleakalā, Maui, Hawaii (USA) |
| A77 | Observatoire Chante-Perdrix - Dauban (France), |
| 098 | Asiago Astrophysical Observatory (Italy) |
| A12 | Stazione Astronomica di Sozzago (Italy) |
| D81 | Yuji Oshima (Japan) |
| A84 | Tubitak National Observatory (Türkiye) |

| Date (UT) | Comet RA (°) | Comet Dec (°) | Measured PA (°) | Computed PA (°) | O-C (°) | Computed angle to sky plane (°) |
|---|---|---|---|---|---|---|
| 2010-01-19 | 192.60 | -3.40 | 77.3 | 74.9 | 2.5 | 12.8 |
| 2010-02-09 | 202.33 | -5.88 | 76.8 | 74.8 | 2.0 | 4.0 |
| 2010-02-10 | 203.32 | -6.08 | 76.0 | 74.7 | 1.3 | 3.1 |
| 2010-02-27 | 208.95 | -6.89 | 76.0 | 74.1 | 1.9 | -2.1 |
| 2010-03-05 | 210.58 | -6.97 | 71.5 | 73.9 | -2.4 | -3.6 |
| 2010-03-20 | 213.21 | -6.64 | 68.7 | 73.6 | -4.9 | -6.2 |
| 2010-03-21 | 213.30 | -6.59 | 65.7 | 73.6 | -7.9 | -6.3 |
| 2010-03-25 | 213.57 | -6.40 | 63.7 | 73.5 | -9.9 | -6.6 |
| 2010-04-17 | 212.73 | -5.10 | 52.7 | 73.8 | -21.1 | -6.2 |
| 2010-04-18 | 212.67 | -5.06 | 52.7 | 73.8 | -21.1 | -6.1 |
| 2010-04-21 | 212.36 | -4.94 | 52.0 | 73.8 | -21.8 | -5.9 |
| 2010-05-07 | 211.02 | -4.74 | n.d. | 74.0 | n.d. | -4.6 |
| 2022-10-30 | 174.67 | 2.94 | 72.0 | 68.3 | 3.7 | 29.3 |
| 2022-11-15 | 185.56 | -1.34 | 70.0 | 72.1 | -2.1 | 20.5 |
| 2022-11-18 | 187.62 | -2.14 | 74.5 | 72.5 | 2.0 | 18.8 |
| 2022-11-23 | 191.75 | -3.74 | 74.5 | 73.1 | 1.4 | 15.3 |
| 2022-11-26 | 193.13 | -4.26 | 77.0 | 73.3 | 3.7 | 14.2 |
| 2022-12-07 | 200.74 | -7.07 | 74.5 | 73.5 | 1.0 | 7.7 |
| 2022-12-31 | 218.00 | -12.63 | 71.0 | 70.6 | 0.4 | -0.5 |
| 2023-01-18 | 229.38 | -15.48 | 61.5 | 66.4 | -4.9 | -16.0 |
| 2023-02-21 | 249.66 | -18.60 | 54.0 | 55.0 | -1.0 | -31.5 |
| 2023-02-22 | 250.18 | -18.64 | 54.5 | 54.6 | -0.1 | -31.9 |
| 2023-02-25 | 251.71 | -18.76 | 52.0 | 53.6 | -1.6 | -33.0 |
| 2023-03-24 | 262.77 | -19.03 | n.d. | 44.7 | n. d | -40.8 |

**Table 3**. The measured PAs of the polar jet are listed for all the observation dates (excluding those when it was not visible), together with the PAs computed according to the estimated positions of the comet's pole, and the relevant O-C. In the last column the computed angle of the spin axis to the sky plane. n.d.=not determined. The RA/Dec coordinates of the comet have been taken from the Horizon/JPL web app Horizons System (nasa.gov).